\documentclass{sf2a-conf2018}
\usepackage{graphicx}
\usepackage{hyperref}
\usepackage[]{natbib}  
\usepackage{epstopdf}
\usepackage{amsmath,amssymb}

\def\BibTeX{{\rm B\kern-.05em{\sc i\kern-.025em b}\kern-.08em
    T\kern-.1667em\lower.7ex\hbox{E}\kern-.125emX}}
\bibpunct{(}{)}{;}{a}{}{,}  

\begin{document}

\TitreGlobal{SF2A 2017}


\title{Consequences of semidiurnal thermal tides \\ on hot Jupiters zonal mean flows}

\runningtitle{Consequences of semidiurnal thermal tides on hot Jupiters zonal mean flows}

\author{P. Auclair-Desrotour}\address{Laboratoire d’Astrophysique de Bordeaux, Univ. Bordeaux, CNRS, B18N, all\'ee Geoffroy Saint-Hilaire, 33615 Pessac, France}



\author{J. Leconte$^1$}

\setcounter{page}{237}


\maketitle


\begin{abstract}
Hot Jupiters are submitted to an intense stellar heating. The resulting thermal tides can torque their atmospheres into asynchronous rotation, while these planets are usually assumed to be locked into spin-orbit synchronization with their host star. Particularly, the thermal atmospheric torque can be greatly enhanced by the dynamical component of the tidal response, that is the component associated with the propagation of internal waves. Owing to the involved complex dynamics, semi-analytical approaches are crucial to understand the physical mechanisms that are responsible for the frequency-resonant behavior of thermal tides, and quantify the atmospheric tidal torque. In this work, we revisit the early works by Arras \& Socrates (2010) and present an improved modeling of thermal tides taking into account rotation and radiative cooling. Using this new modeling, we compute analytically the atmospheric tidal response of hot Jupiters and show that resonances associated with low-frequency internal gravity waves are able to drive asynchronous zonal flows in the range 1-30 days. 
\end{abstract}



\section{Introduction}

Understanding the general circulation of hot Jupiters is crucial to constrain observationally their properties (temperature, structure, day-night heat transport, circulation regimes). Particularly, it appears as a key step to relate these properties with the Doppler shift in transmission spectra that can be measured in orbital phase curves of secondary eclipses \citep[e.g.][]{Rauscher2014}. Because they orbit very close to their host stars, hot Jupiters undergo strong gravitational tides. Therefore, they are expected to be torqued towards spin-orbit synchronization, that is the equilibrium state where the planet rotation rate exactly equalizes the orbital frequency, over short timescales \citep[typically a few millions years, see][]{SG2002,OL2004,Showman2015}. Hence, tidal forces should lock a planet into synchronous rotation before they circularize its orbit \citep[e.g.][]{Rasio1996}. 

However, other mechanisms can compensate the effects of gravitational tides and force a fast super-rotating equatorial jet in the atmosphere of the planet, which is also submitted to a strong day-night heating contrast. One may invoke for instance the non-linear equatorward transport of angular momentum by Rossby waves \citep[e.g.][]{SP2011,Showman2015}, or thermal tides for their ability to induce a tidal torque in opposition with that induced by gravitational tides \citep[][]{GO2009,AS2010}. This latter effect has been particularly studied in the case of terrestrial planets such as Venus, which is locked into the observed asynchronous rotation by a competition between gravitationally and thermally induced tidal torques \citep[][]{GS1969,ID1978,DI1980,CL2001,Leconte2015,ADLM2017a}. Its efficiency is based in this case upon the fact that the solid part of the planet can support the surface load resulting from variations of the atmospheric mass distribution with negligible distortions, thus leading to a net quadrupole. 

Since there is not solid part at the base of the atmosphere in hot Jupiters, the lower regions of the planet tend to hydrostatically compensate the variation of atmospheric mass distribution in the low-frequency range, preventing thereby a net quadrupole to form. Nevertheless, by proceeding to a linear analysis of the atmospheric tidal response derived from the classical tidal theory \citep{CL70}, \cite{AS2010} showed evidence of the important role played by the so-called ``dynamical tide'' \citep[][]{Zahn1966a}, that is the component of the tidal response resulting from the propagation of internal waves, which can lead to an energy dissipation enhanced by several orders of magnitude.

\cite{AS2010} computed the tidal torque exerted on the atmosphere in the absence of rotation and dissipative processes, which provided a diagnosis about the impact of dynamical tide on the tidal torque but let aside potentially important physical ingredients. Therefore, we revisit here their early study by introducing in a self-consistent way in the atmospheric tidal response the effects of radiative cooling and rotation. We thus characterize the wave-like structure of the atmospheric tidal response, examine the evolution of the tidal torque with the forcing frequency, and quantify its dependence on rotation and radiative cooling. Results summarized in the following are detailed in \cite{ADL2018}. 
  
\section{Structure of the tidal response}
\label{sec:structure}

\begin{figure}[ht!]
 \centering
 \includegraphics[width=0.48\textwidth,clip]{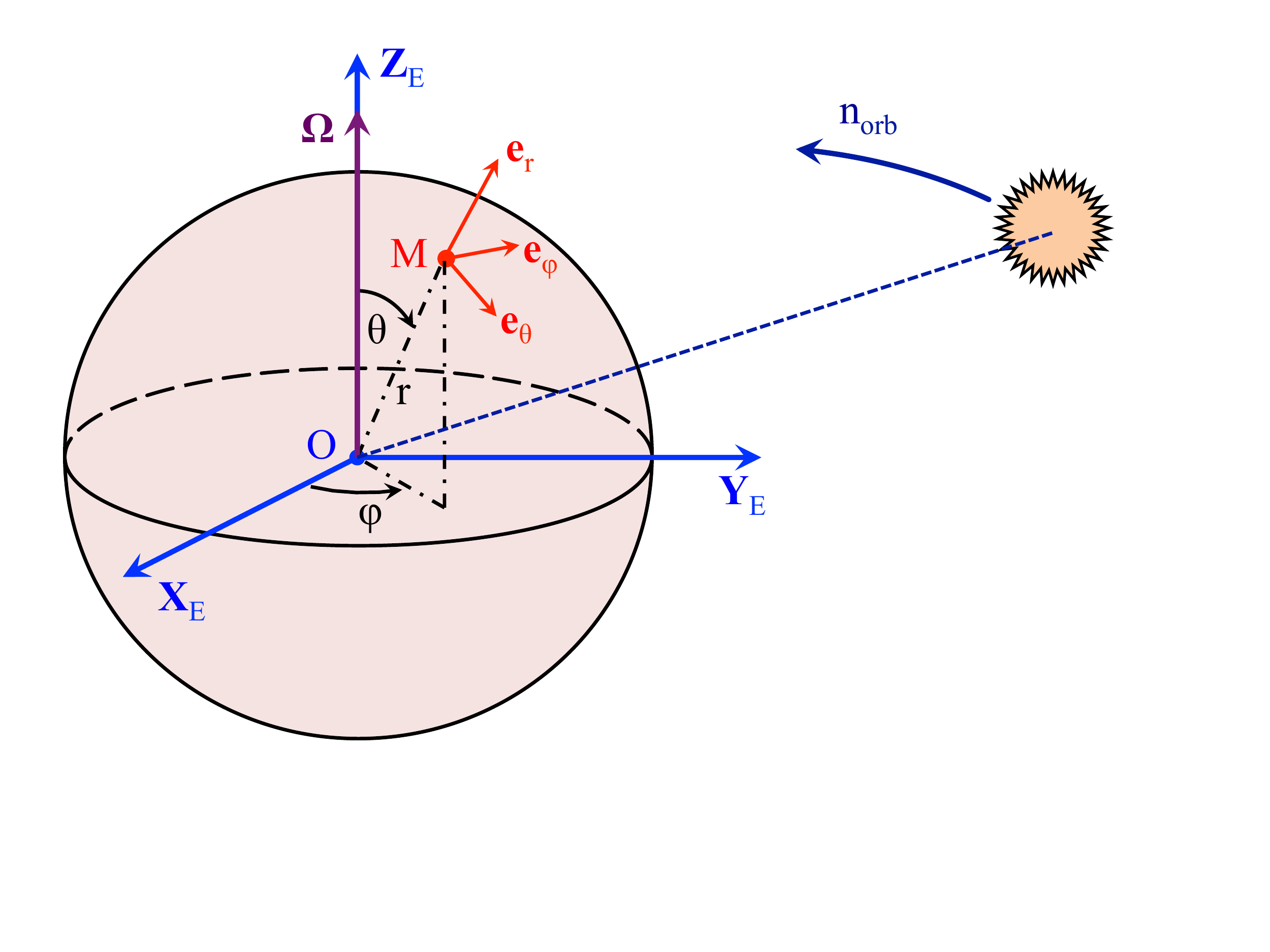}%
 \includegraphics[width=0.48\textwidth,trim =  1.5cm 1.8cm 1.5cm 1.5cm,clip]{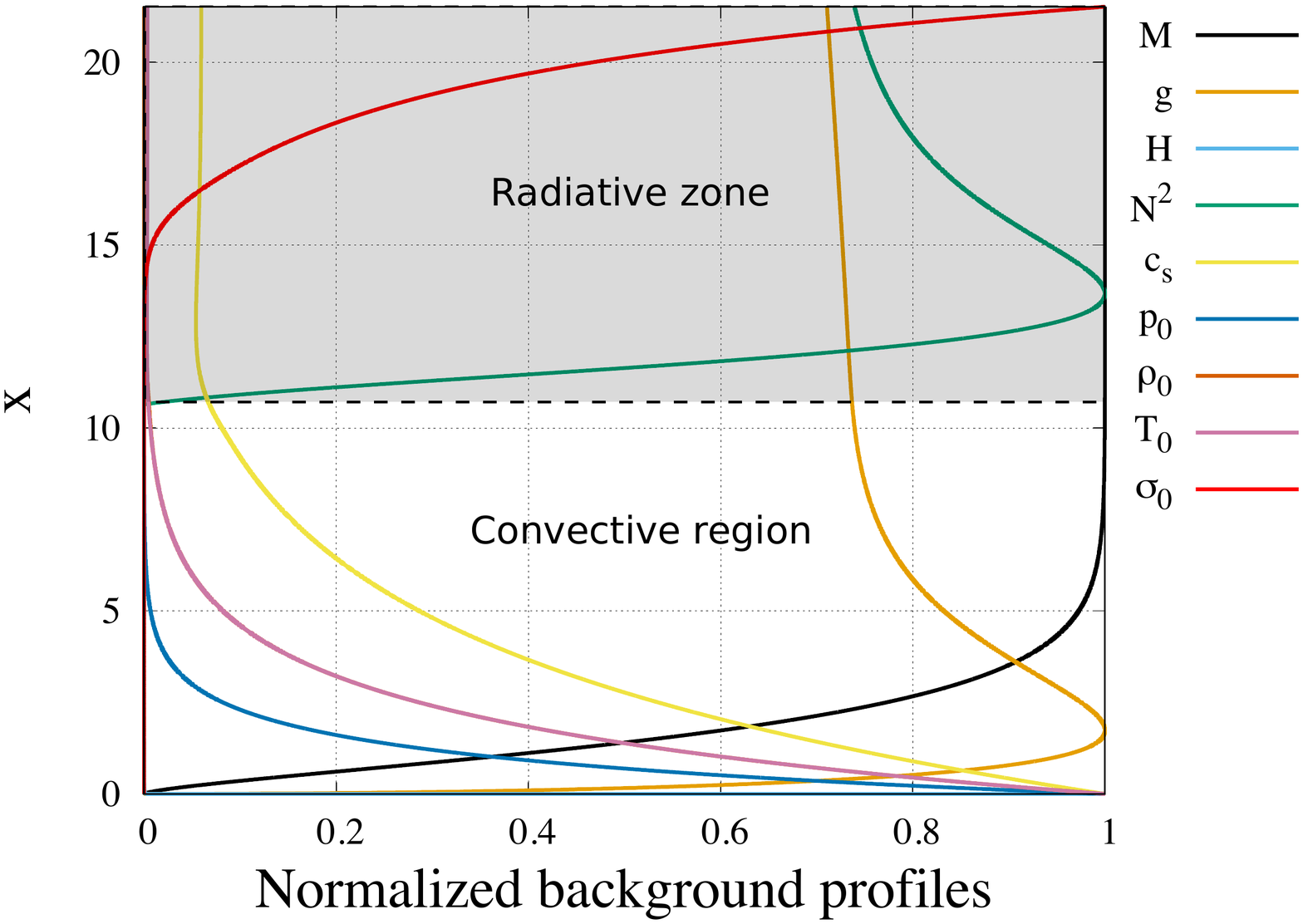}      
  \caption{{\bf Left:} Reference frames and systems of coordinates. The notations $\boldsymbol{\Omega}$ and $n_{\rm orb}$ designate the rotation vector and the orbital angular velocity, respectively. {\bf Right:} Background vertical profiles normalized by their maxima as functions of the pressure altitude $x$.}
  \label{auclair-desrotour2:fig1}
\end{figure}

Following along the line by \cite{AS2010}, we consider a fluid spherical planet of radius $R_{\rm p}$ and mass $M_{\rm p}$ orbiting circularly its host star, of mass $M_\star$ (see Fig.~\ref{auclair-desrotour2:fig1}, left panel). The planet spin is defined by the angular velocity $\Omega$ and its mean motion by the orbital frequency $n_{\rm orb}$. We adopt for background distributions the vertical profiles used by \cite{AS2010}. As shown by Fig.~\ref{auclair-desrotour2:fig1}, these profiles define a bi-layered planet composed of a thin atmosphere, which is stably stratified with respect to convection, and a thick convective region. Background distributions of gravity $g$, pressure $p_0$, density $\rho_0$ and temperature $T_0$ are supposed to be functions of the radial coordinate $r$ only, and mean flows are ignored. 

The planet undergoes both the tidal gravitational and thermal forcings resulting from the tidal gravitational potential of the star and the absorption of the incoming stellar flux, respectively. However, we focus here on the thermal component, as it only affects the thin stably stratified atmosphere, where approximations allowing the mathematical treatment of the classical tidal theory can be assumed (typically the traditional approximation). The stellar heating generates a tidal response, which is considered as a small perturbation in the vicinity of a given state of equilibrium in the linear analysis adopted for this work.  In the absence of obliquity, the tidal torque exerted on the planet is generated by the semidiurnal tide only, of frequency $\sigma = 2 \left( \Omega - n_{\rm orb} \right)$. It can be written 

\begin{equation}
\begin{array}{rcl}
\displaystyle \mathcal{T} = 2 \pi \sqrt{\frac{3}{5}} \left( \frac{M_\star}{M_\star + M_{\rm p}} \right) n_{\rm orb}^2 \Im \left\{ Q_2^{2,\sigma} \right\}, & \mbox{with} & \displaystyle Q_2^{2,\sigma} = \sum_n A_{n,2}^{2,\nu} \int_0^{+ \infty} r^4 \delta \rho_n^{2,\sigma } dr. 
\end{array}
\end{equation}
 
In the above expression, we have introduced the quadrupole moment $Q_2^{2,\sigma}$, which describes the global variation of mass distribution associated by the tidal bulge. Through Coriolis effects, the rotation couples spatial distributions of forcings to Hough modes \citep[][]{CL70,LS1997}, designed by the subscript $n$. Therefore, $Q_2^{2,\sigma}$ is the sum of contributions of modes weighted by the coupling coefficients $A_{n,2}^{2,\nu}$. The associated density variations are denoted $\delta \rho_n^{2,\sigma}$ while $r$ and $dr$ designate the radial coordinate and its differential, respectively. 

\begin{figure*}
   \centering
   \includegraphics[height=0.4cm]{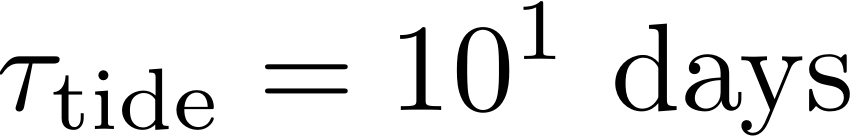} \hspace{0.8cm}
   \includegraphics[height=0.4cm]{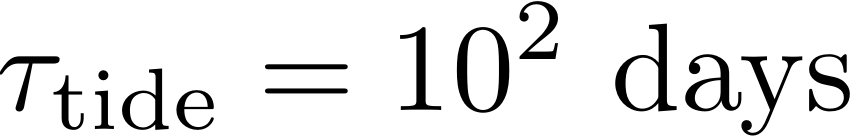} \hspace{0.8cm} \hspace{0.5cm}
   \includegraphics[height=0.4cm]{auclair-desrotour2_fig2a.pdf} \hspace{0.8cm}
   \includegraphics[height=0.4cm]{auclair-desrotour2_fig2b.pdf} \hspace{0.8cm}\\
   \raisebox{1.0cm}{\includegraphics[width=0.02\textwidth]{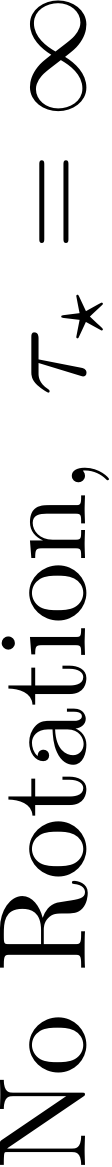}} \hspace{0.1cm}
   \raisebox{1.5\height}{\includegraphics[width=0.015\textwidth]{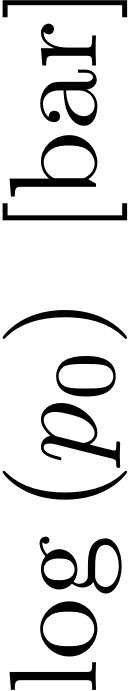}}
   \includegraphics[width=0.20\textwidth,trim = 2.5cm 2.2cm 3.cm 0.cm, clip]{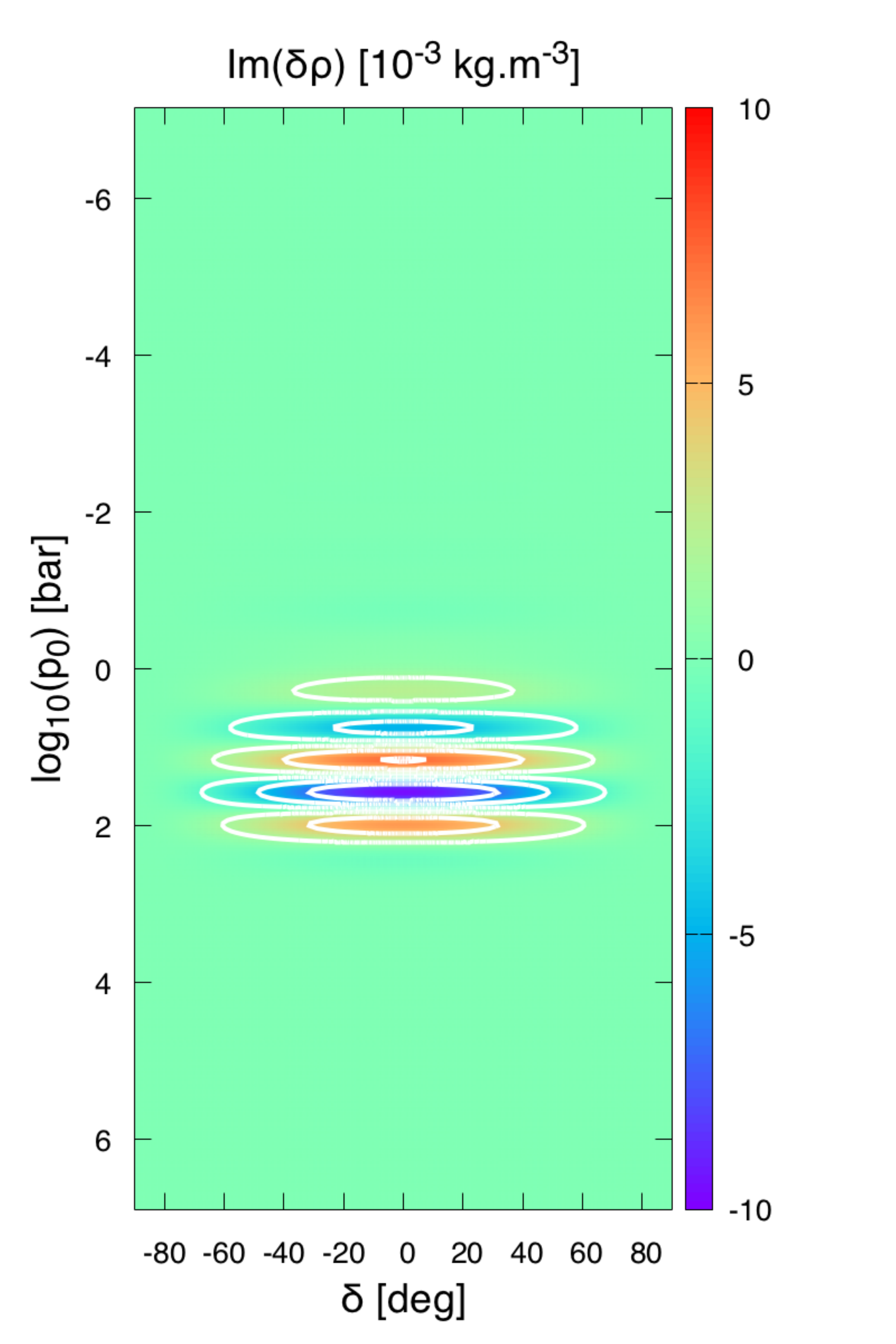}
   \includegraphics[width=0.20\textwidth,trim = 2.5cm 2.2cm 3.cm 0.cm, clip]{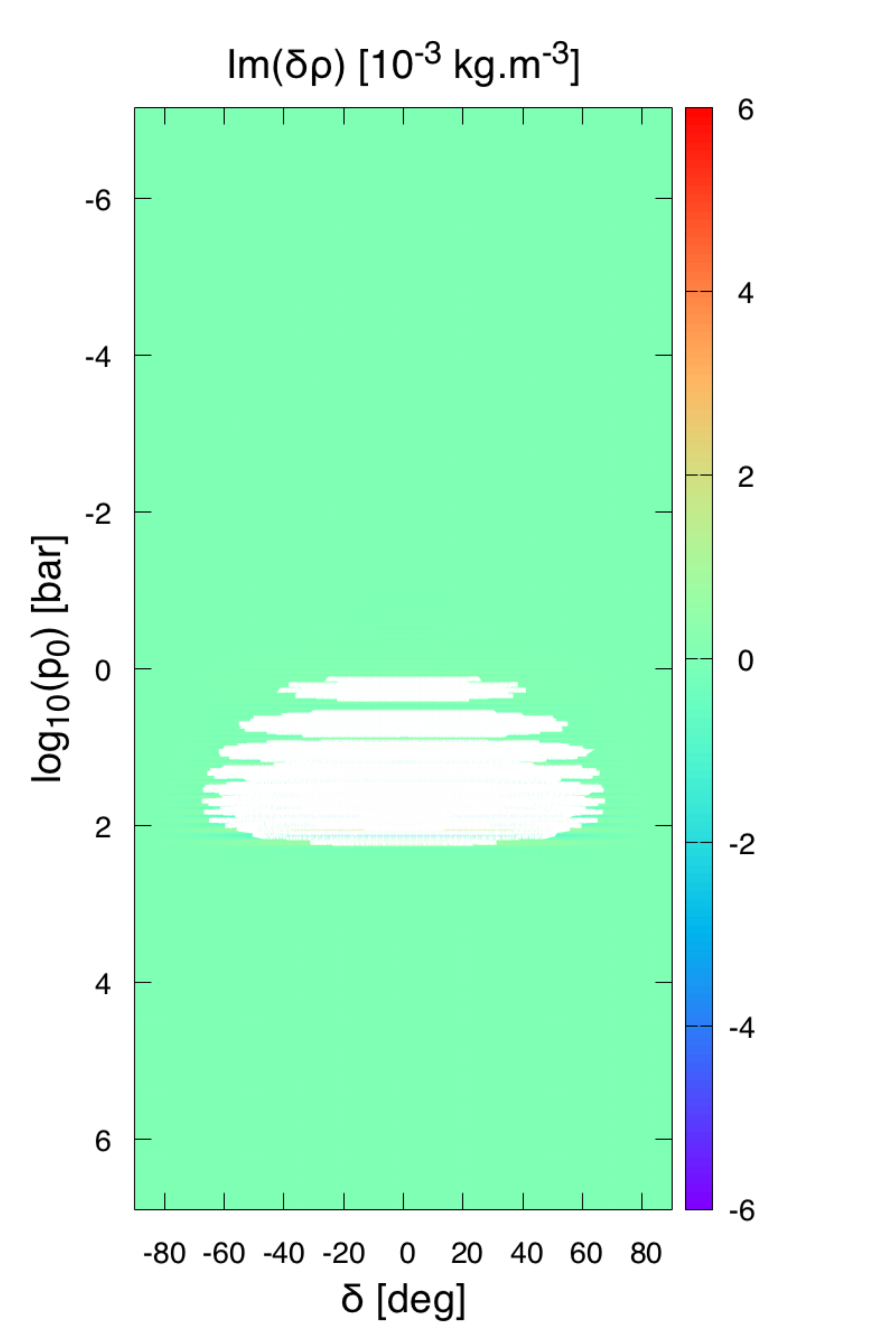} \hspace{0.5cm}
   \includegraphics[width=0.20\textwidth,trim = 2.5cm 2.2cm 3.cm 0.cm, clip]{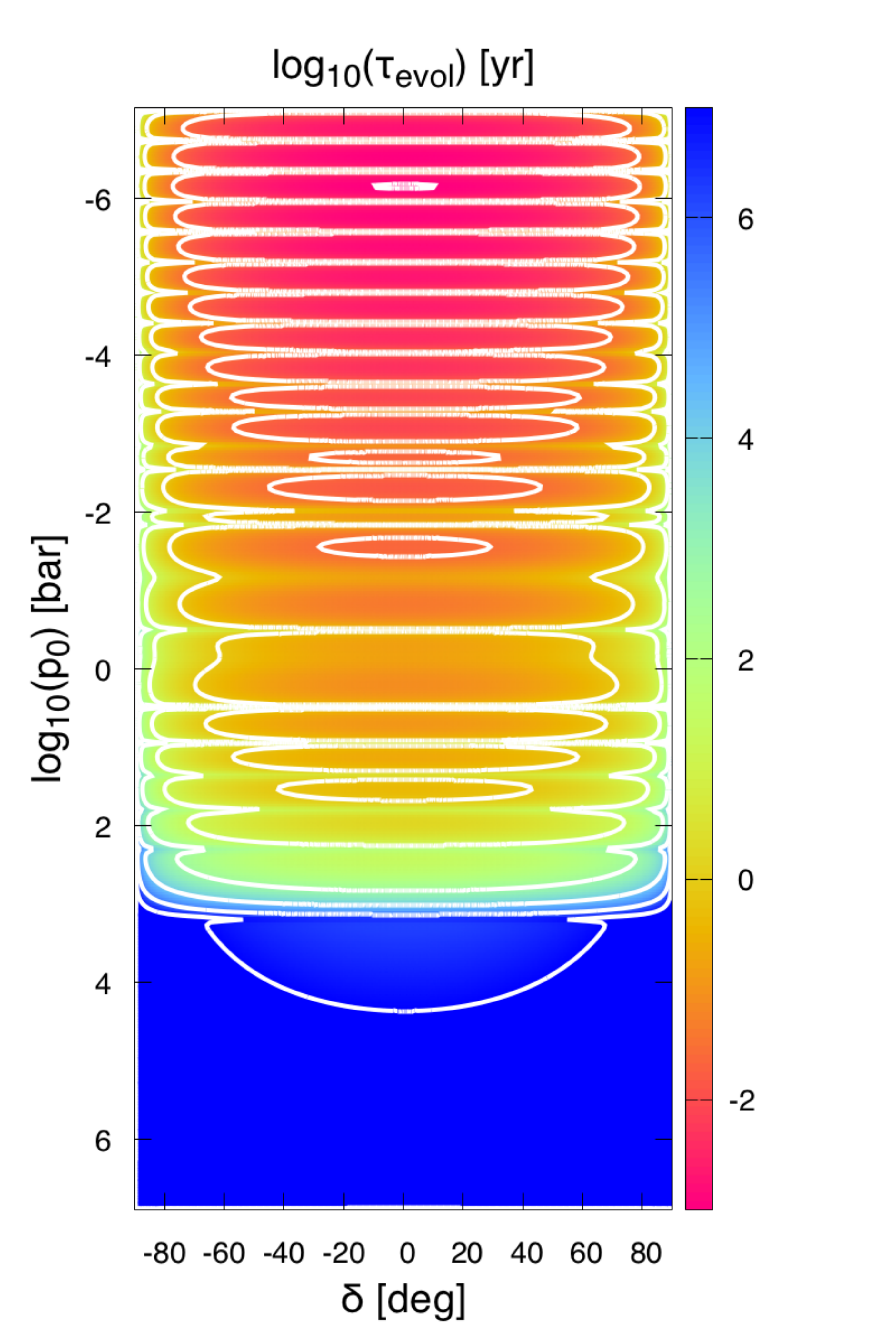} 
   \includegraphics[width=0.20\textwidth,trim = 2.5cm 2.2cm 3.cm 0.cm, clip]{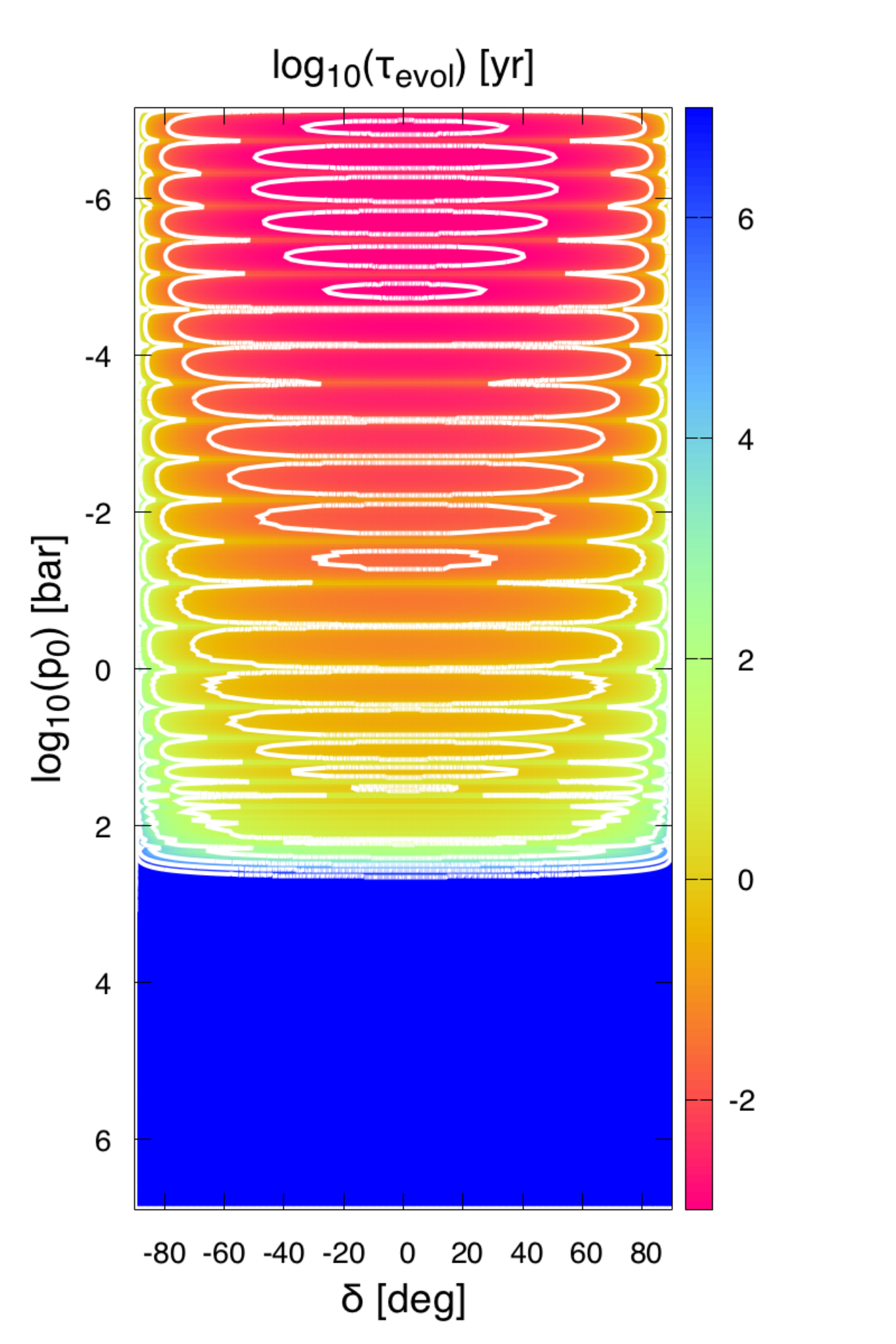} \\
   \raisebox{1.0cm}{\includegraphics[width=0.02\textwidth]{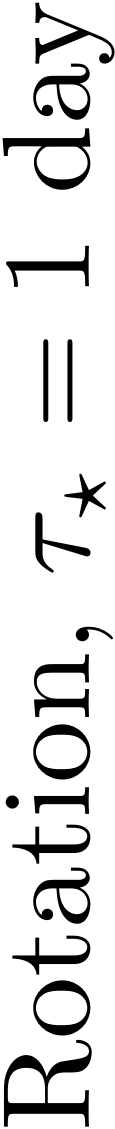}} \hspace{0.1cm}
   \raisebox{1.5\height}{\includegraphics[width=0.015\textwidth]{auclair-desrotour2_fig2d.pdf}}
   \includegraphics[width=0.20\textwidth,trim = 2.5cm 2.2cm 3.cm 0.cm, clip]{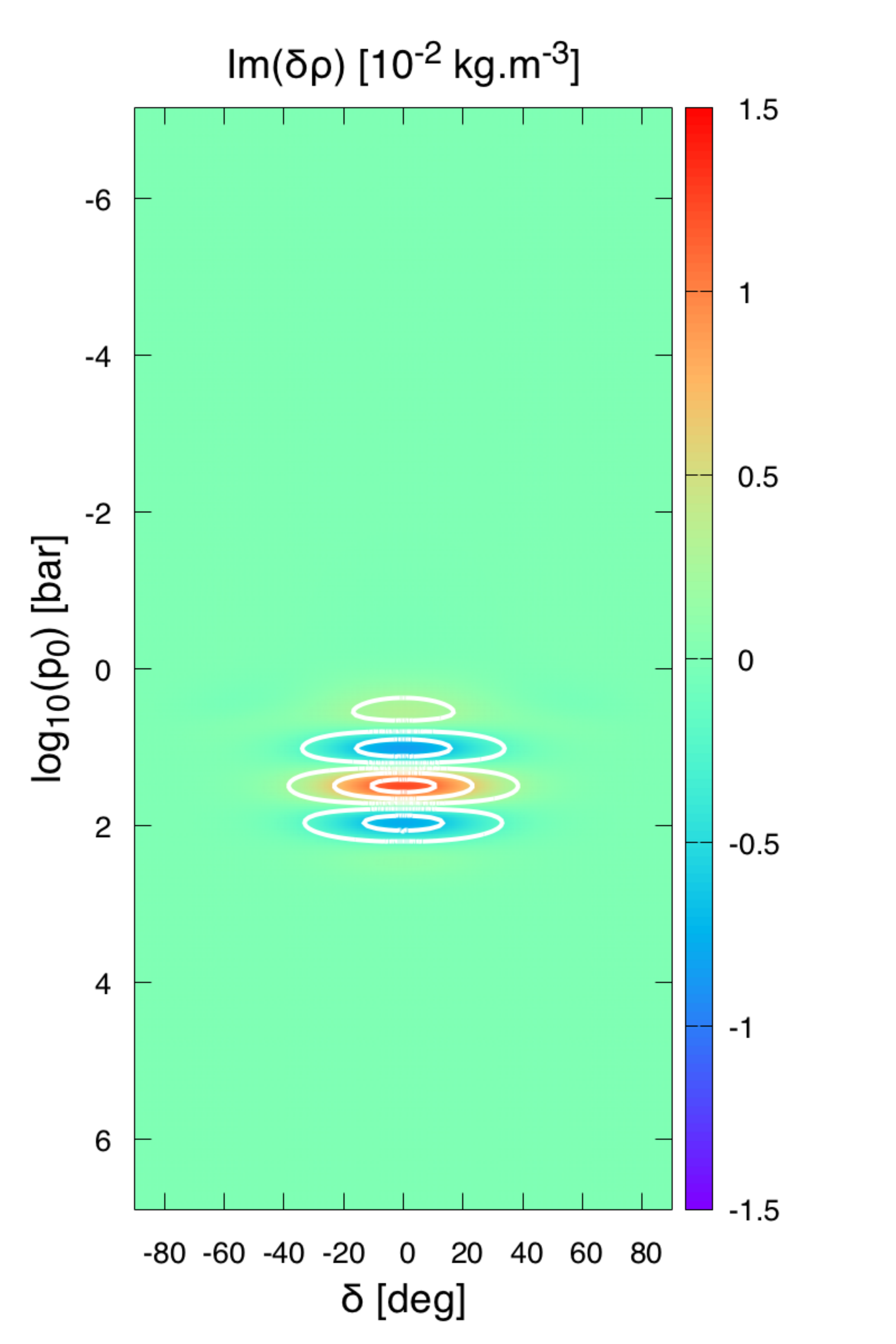}
   \includegraphics[width=0.20\textwidth,trim = 2.5cm 2.2cm 3.cm 0.cm, clip]{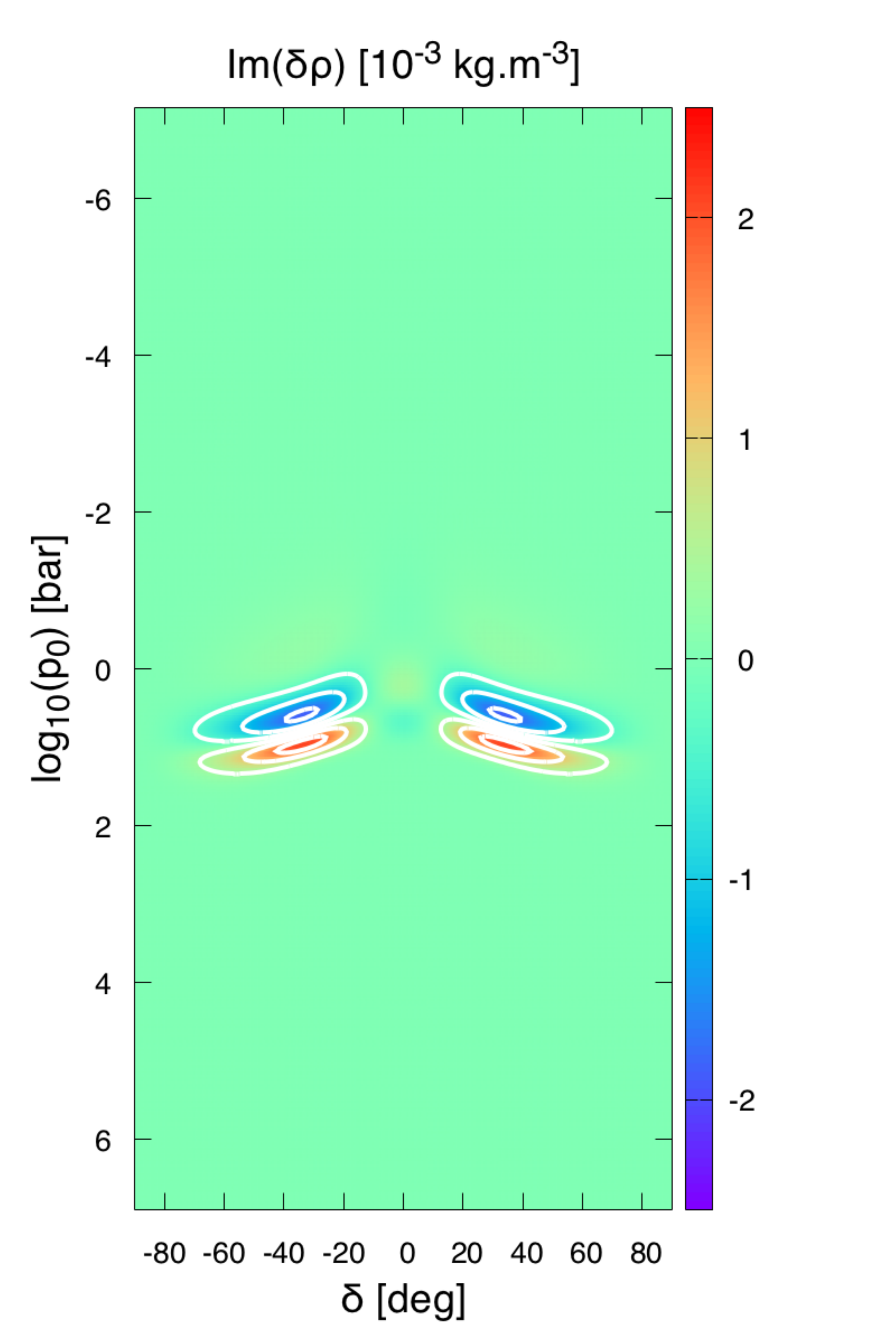}\hspace{0.5cm}
   \includegraphics[width=0.20\textwidth,trim = 2.5cm 2.2cm 3.cm 0.cm, clip]{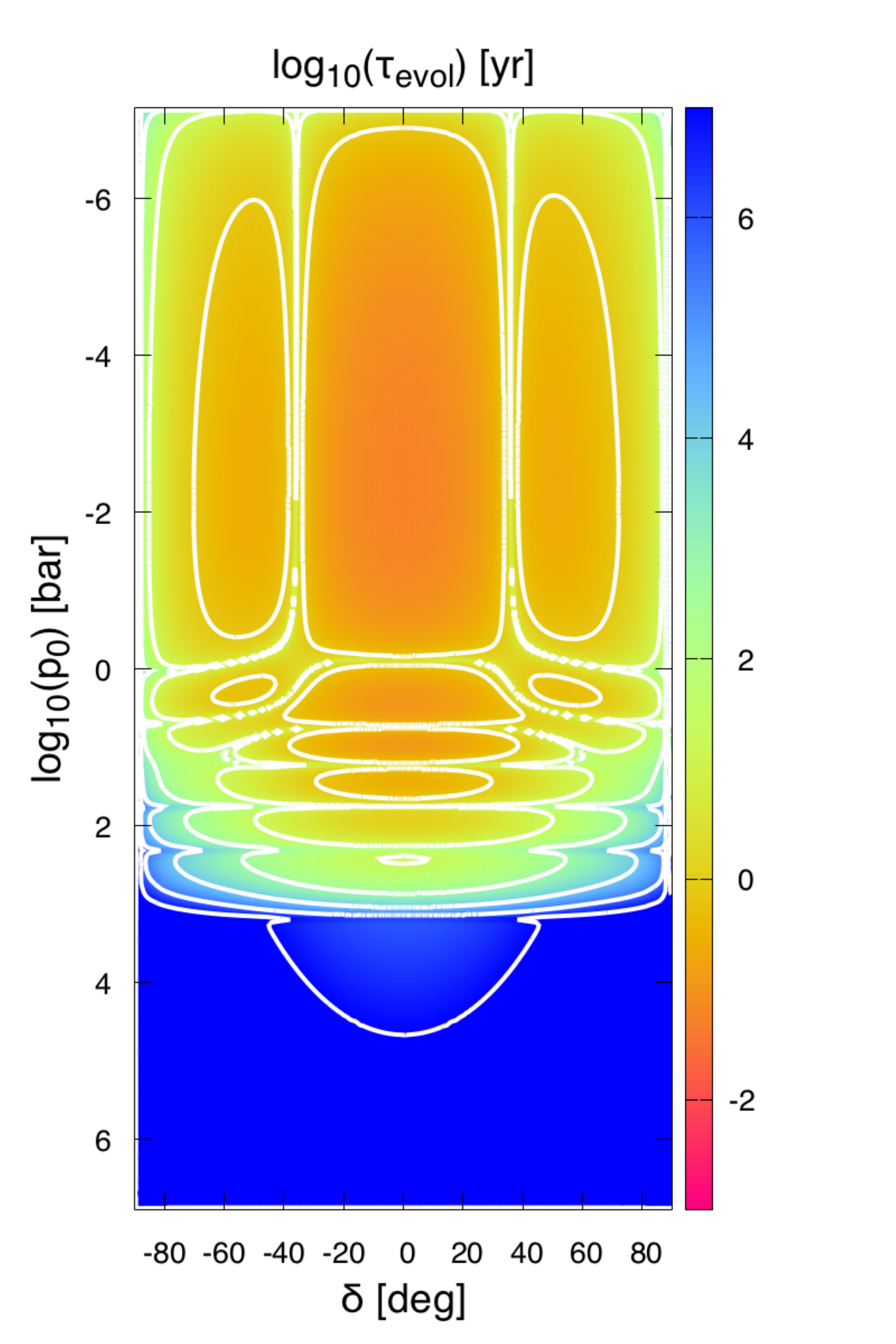} 
   \includegraphics[width=0.20\textwidth,trim = 2.5cm 2.2cm 3.cm 0.cm, clip]{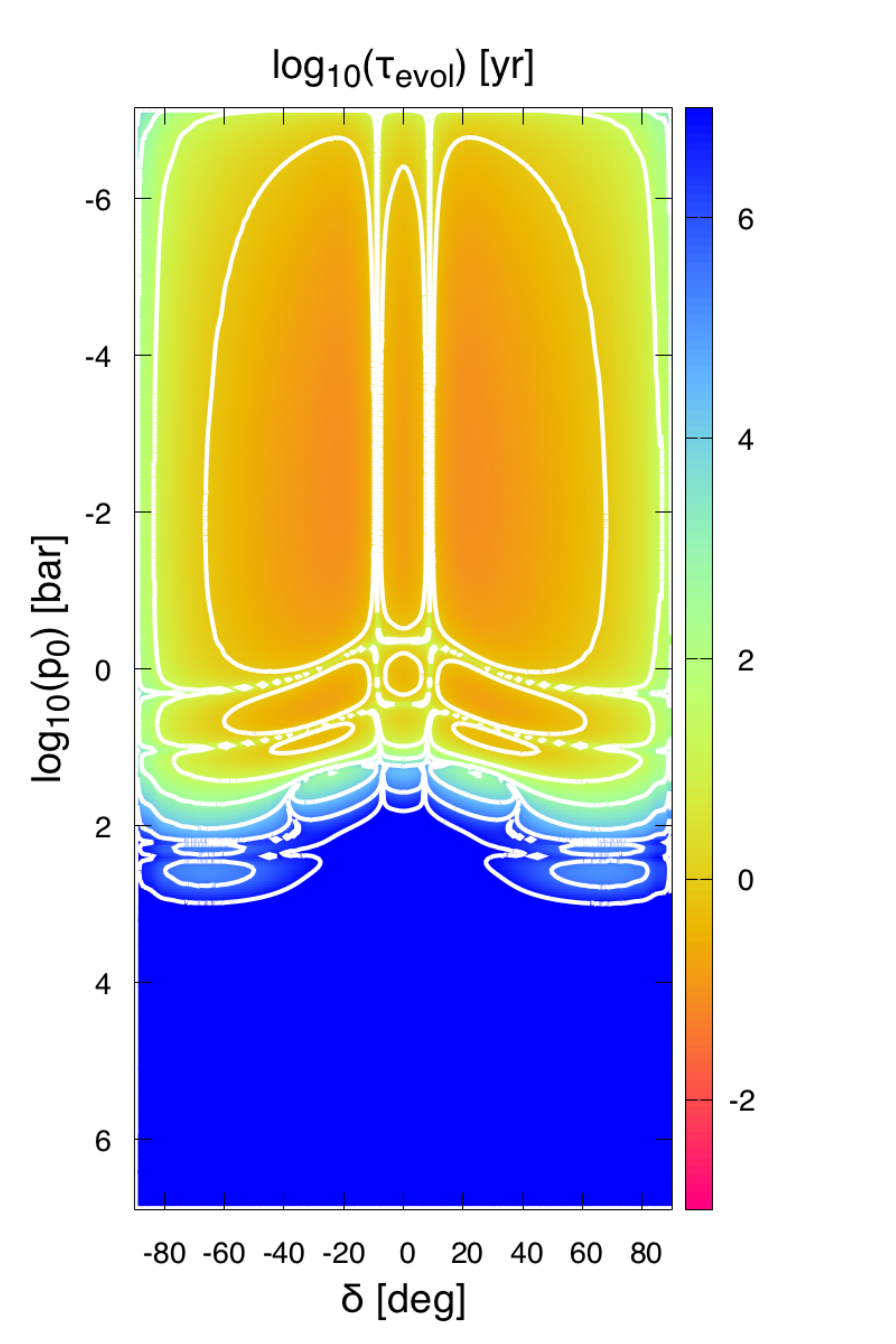}  \\
   \hspace{1.9cm}
   \hspace{0.025\textwidth}
   \includegraphics[height=0.3cm]{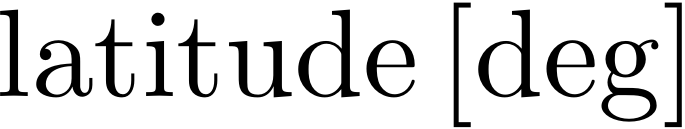} \hspace{1.6cm}
   \includegraphics[height=0.3cm]{auclair-desrotour2_fig2p.pdf} \hspace{1.6cm} \hspace{0.5cm}
   \includegraphics[height=0.3cm]{auclair-desrotour2_fig2p.pdf} \hspace{1.6cm}
   \includegraphics[height=0.3cm]{auclair-desrotour2_fig2p.pdf} \hspace{1.6cm}
   \caption{Structure of the atmospheric tidal response to the semidiurnal thermal forcing for tidal periods $\tau_{\rm tide} = 10$ and 100 days. Quantities are plotted as functions of latitude (horizontal axis, degrees) and pressure levels in logarithmic scale (vertical axis, bars). {\bf Left:} Imaginary part of density fluctuations. {\bf Right:} Characteristic timescale necessary for the semidiurnal tide to generate a zonal jet of velocity $V_{\rm jet} = 1 \ {\rm km.s^{-1}} $ in logarithmic scale. {\bf Top:} Static planet (rotation ignored) with no dissipative processes. {\bf Bottom:} Rotating planet with radiative cooling. }
       \label{auclair-desrotour2:fig2}%
\end{figure*}

By solving the Laplace's tidal equation \citep[e.g.][Eq.~(51)]{LC1969}, we obtain the $A_{n,2}^{2,\nu}$ and the eigenvalues associated with Hough modes. These laters are then used to integrate numerically the equation describing the vertical structure of the tidal response and compute the associated density variations. We also introduce the characteristic timescale $\tau_{\rm evol}$, which is the timescale necessary for the force induced by the local variation of mass distribution to generate a zonal jet of $1 \ {\rm km.s^{-1}}$ assuming a constant acceleration. In order to give an overview of the obtained results, the imaginary part of the quadrupolar density anomaly and the corresponding evolution timescale are plotted in Fig.~\ref{auclair-desrotour2:fig2} as functions of the latitude and pressure levels in logarithmic scale. Results are given for the tidal periods $\tau_{\rm tide} = 2 \pi / \left| \sigma \right| = 10,100$~days and two cases:  the static adiabatic case treated by \cite{AS2010} and the case of a rotating planet with a radiative timescale $\tau_\star = 1$~day. 

Figure~\ref{auclair-desrotour2:fig2} illustrates how radiative cooling and rotation affect the structure of the tidal response. The observed oscillatory behaviour (top panels) is due to internal gravity waves forming the dynamical tide, which can propagate in the radiative atmosphere. In the absence of dissipation, the vertical wavelength of these waves tends to zero when the tidal period increases, i.e. when the planet tends to synchronization. The introduction of radiative cooling leads to a more realistic behaviour, oscillations being damped for $\tau_{\rm tide} \gg \tau_\star$. As regards the impact of rotation, we retrieve the well-known equatorial trapping of gravity waves for $\tau_{\rm tide} = 10$~days. In the vicinity of synchronization, $\left| \nu \right| > 1$. The distribution of tidal heating is thus coupled with Rossby waves, which leads to the particular pattern observed for $\tau_{\rm tide}  =  100$~days. In all cases, Fig.~\ref{auclair-desrotour2:fig2} shows a strong decoupling between the atmosphere and the convective region. The tidal force induced by the thermal forcing affect the irradiated zone only, gravity waves being unable to propagate in the convective region.

\section{Evolution of the atmospheric tidal torque with the tidal frequency}

\begin{figure}[ht!]
 \centering
 \includegraphics[width=0.48\textwidth,clip]{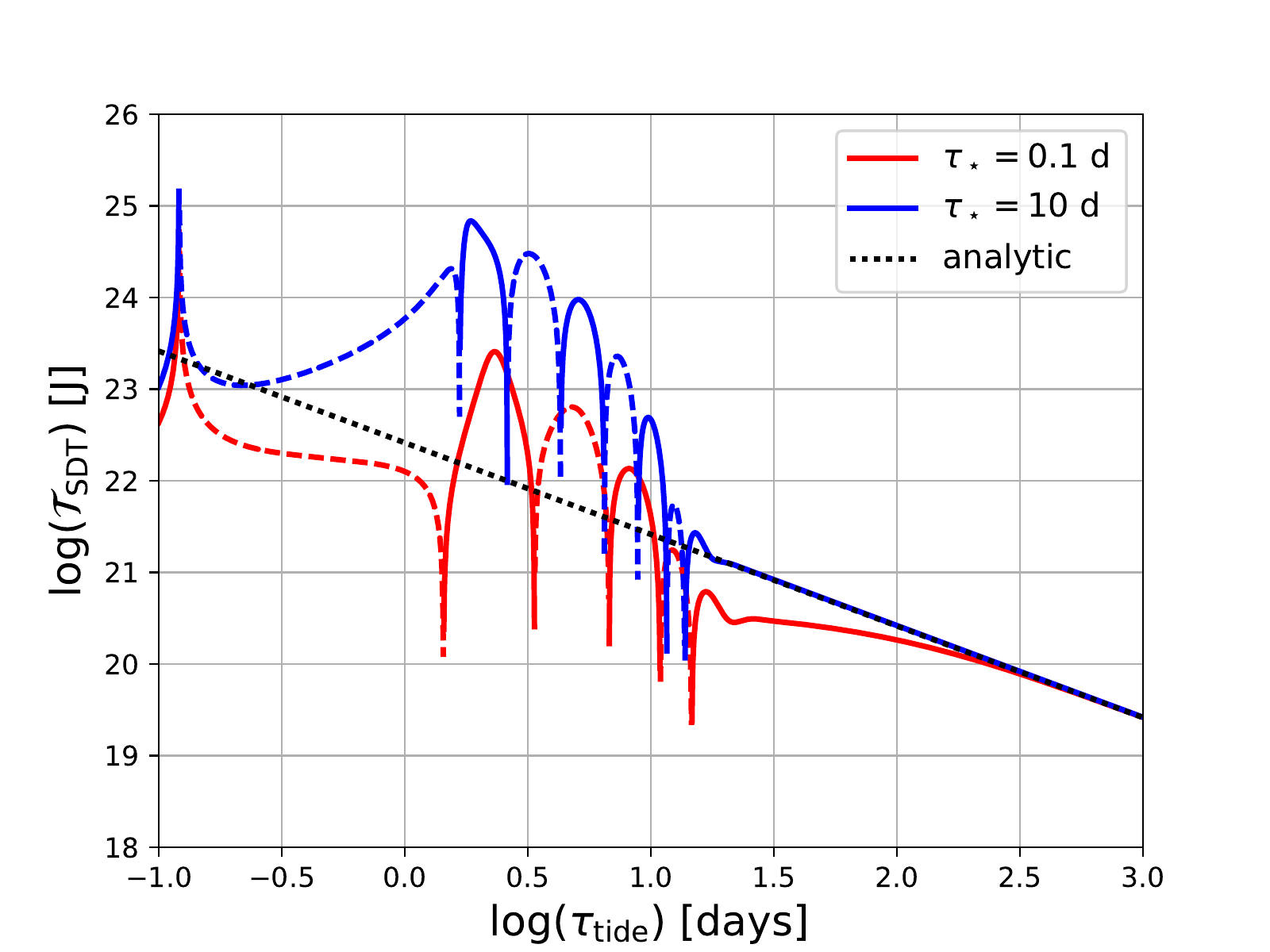}%
 \includegraphics[width=0.48\textwidth,clip]{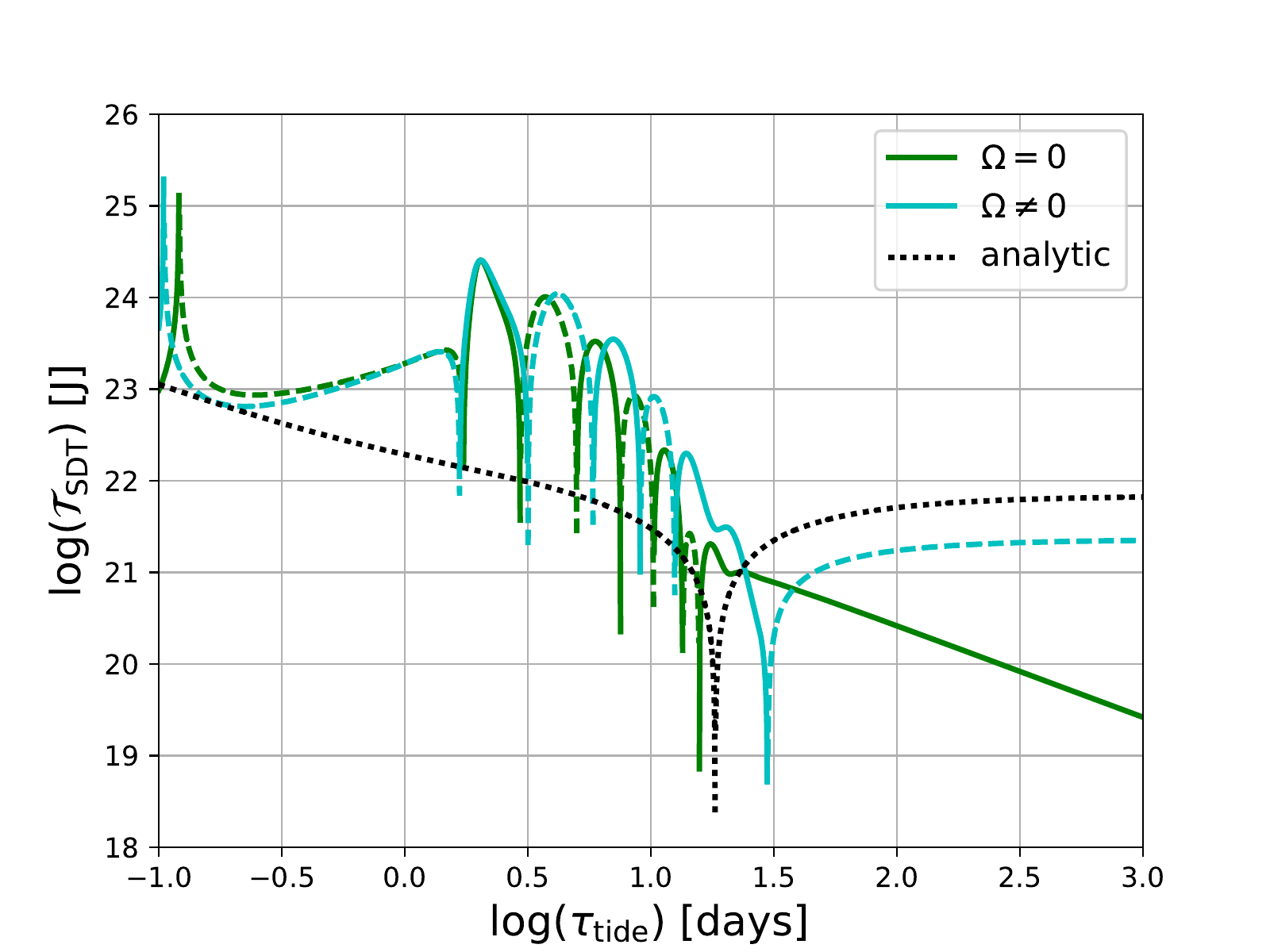}      
  \caption{Atmospheric tidal torque created by the semidiurnal thermal tide (J) as a function of the tidal period (days) in logarithmic scales.  {\bf Left:} Static planet with characteristic timescales of radiative cooling $\tau_\star = 0.1$~days (red line) and $\tau_\star = 10$~days (blue line). {\bf Right:} Static (green line) and rotating (cyan line) planet with $\tau_\star = 1$~day. In both panels, solid (dashed) lines correspond to negative (positive) torques, pushing the planet away from (toward) spin-orbit synchronous rotation. The black dotted line designates the equilibrium tidal torque derived analytically in the low-frequency range. }
  \label{auclair-desrotour2:fig3}
\end{figure}

We now compute the variation of the total tidal torque exerted on the atmosphere with the tidal frequency. A first analysis in the zero-frequency limit leads to the analytical expression of the quadrupole moment 

 \begin{equation}
 Q_2^{2,\sigma} \approx \sum_n A_{n,2}^{2,\nu} \left[ 1 - \frac{30}{\Lambda_n^{2,\nu}} \right] \int_0^{+\infty} \rho_0 r^4  \left( \frac{\sigma^2}{N^2} \right) \frac{J_n}{i \sigma T_0 C_p},
 \label{Qana}
 \end{equation}

\noindent where $i$ designates the imaginary number, $C_p$ the thermal capacity of the gas per unit mass, $N$ the Brunt-V\"ais\"al\"a frequency, supposed constant in the isothermal approximation, and $J_n$ the thermal forcing per unit mass resulting from the absorption of the incoming stellar flux. This formula generalizes that obtained by \cite{AS2010} in the static adiabatic case. The total torque created by the semidiurnal tide is plotted in Fig.~\ref{auclair-desrotour2:fig3} as a function of the tidal period ($\tau_{\rm tide} = 2 \pi / \left| \sigma \right|$) in logarithmic scales, for two different radiative time scales (left panel), and for the static and rotating cases (right panel). In both panels, the analytic formula given by Eq.~(\ref{Qana}) is designated by the dotted black line.

In the range $1-30$~days, we recover the resonant behaviour previously obtained by \cite{AS2010}. Resonances are due to the wave-like component of the tidal response highlighted by Fig.~\ref{auclair-desrotour2:fig2}. They can enhance the tidal torque by several orders of magnitudes, this effect being stronger in the absence of damping. Although rotation slightly modifies resonances, it mainly affects the tidal response in the low-frequency regime because of the coupling of the forcing with Hough modes discussed in Sect.~\ref{sec:structure}. However, note that this behaviour is not realistic in the zero-frequency limit since other dissipative processes like friction will annihilate the coupling induced by Coriolis effects beyond a given tidal period. This processes will make the torque converge to the linear behaviour obtained in the static case.

\section{Conclusion}

Motivated by the understanding of the role played by thermal atmospheric tides in the general circulation of hot Jupiters, we studied the ability of the induced tidal torque to generate strong asynchronous jets. To do that, we revisited the early work by \cite{AS2010} by introducing Coriolis effects and radiative cooling in the dynamics of thermal tides. We computed the 3D structure of the tidal response numerically in the general case using the linear analysis of the classical theory of atmospheric tides. This provided us a diagnosis about the way the atmosphere is affected by tidally induced variations of mass distribution. We recovered in the static adiabatic case resonances identified before by \cite{AS2010}. These resonances are associated with the dynamical tide and result from the propagation of internal gravity waves. They can increase the total tidal torque exerted on the atmosphere by several orders of magnitude. However, they are attenuated by radiative cooling. Coriolis effects mainly change the asymptotic behaviour of the torque in the low-frequency range. In spite of the limitations inherent to the linear analysis, this approach enables a full characterization of the thermally-induced tidal torque undergone by the atmosphere of the planet as a function of the tidal frequency and physical parameters of the system. It may be consolidated in further works by the study of the impact of friction on the atmospheric tidal response in the zero-frequency limit.

\begin{acknowledgements}
The authors acknowledge funding by the European Research Council through the ERC grant WHIPLASH 679030.
\end{acknowledgements}

\bibliographystyle{aa}  
\bibliography{auclair-desrotour2} 

\begin{thebibliography}{19}
\expandafter\ifx\csname natexlab\endcsname\relax\def\natexlab#1{#1}\fi

\bibitem[{{Arras} \& {Socrates}(2010)}]{AS2010}
{Arras}, P. \& {Socrates}, A. 2010, \apj, 714, 1

\bibitem[{{Auclair-Desrotour} {et~al.}(2017){Auclair-Desrotour}, {Laskar}, \&
  {Mathis}}]{ADLM2017a}
{Auclair-Desrotour}, P., {Laskar}, J., \& {Mathis}, S. 2017, \aap, 603, A107

\bibitem[{{Auclair-Desrotour} \& {Leconte}(2018)}]{ADL2018}
{Auclair-Desrotour}, P. \& {Leconte}, J. 2018, \aap, 613, A45

\bibitem[{{Chapman} \& {Lindzen}(1970)}]{CL70}
{Chapman}, S. \& {Lindzen}, R. 1970, {Atmospheric tides: Thermal and
  gravitational. Reidel.}

\bibitem[{{Correia} \& {Laskar}(2001)}]{CL2001}
{Correia}, A.~C.~M. \& {Laskar}, J. 2001, \nat, 411, 767

\bibitem[{{Dobrovolskis} \& {Ingersoll}(1980)}]{DI1980}
{Dobrovolskis}, A.~R. \& {Ingersoll}, A.~P. 1980, \icarus, 41, 1

\bibitem[{{Gold} \& {Soter}(1969)}]{GS1969}
{Gold}, T. \& {Soter}, S. 1969, \icarus, 11, 356

\bibitem[{{Gu} \& {Ogilvie}(2009)}]{GO2009}
{Gu}, P.-G. \& {Ogilvie}, G.~I. 2009, \mnras, 395, 422

\bibitem[{{Ingersoll} \& {Dobrovolskis}(1978)}]{ID1978}
{Ingersoll}, A.~P. \& {Dobrovolskis}, A.~R. 1978, \nat, 275, 37

\bibitem[{{Leconte} {et~al.}(2015){Leconte}, {Wu}, {Menou}, \&
  {Murray}}]{Leconte2015}
{Leconte}, J., {Wu}, H., {Menou}, K., \& {Murray}, N. 2015, Science, 347, 632

\bibitem[{{Lee} \& {Saio}(1997)}]{LS1997}
{Lee}, U. \& {Saio}, H. 1997, \apj, 491, 839

\bibitem[{{Lindzen} \& {Chapman}(1969)}]{LC1969}
{Lindzen}, R.~S. \& {Chapman}, S. 1969, \ssr, 10, 3

\bibitem[{{Ogilvie} \& {Lin}(2004)}]{OL2004}
{Ogilvie}, G.~I. \& {Lin}, D.~N.~C. 2004, \apj, 610, 477

\bibitem[{{Rasio} {et~al.}(1996){Rasio}, {Tout}, {Lubow}, \&
  {Livio}}]{Rasio1996}
{Rasio}, F.~A., {Tout}, C.~A., {Lubow}, S.~H., \& {Livio}, M. 1996, \apj, 470,
  1187

\bibitem[{{Rauscher} \& {Kempton}(2014)}]{Rauscher2014}
{Rauscher}, E. \& {Kempton}, E.~M.~R. 2014, \apj, 790, 79

\bibitem[{{Showman} \& {Guillot}(2002)}]{SG2002}
{Showman}, A.~P. \& {Guillot}, T. 2002, \aap, 385, 166

\bibitem[{{Showman} {et~al.}(2015){Showman}, {Lewis}, \&
  {Fortney}}]{Showman2015}
{Showman}, A.~P., {Lewis}, N.~K., \& {Fortney}, J.~J. 2015, \apj, 801, 95

\bibitem[{{Showman} \& {Polvani}(2011)}]{SP2011}
{Showman}, A.~P. \& {Polvani}, L.~M. 2011, \apj, 738, 71

\bibitem[{{Zahn}(1966)}]{Zahn1966a}
{Zahn}, J.~P. 1966, Annales d'Astrophysique, 29, 313

\end{thebibliography}

\end{document}